\documentclass[12pt]{article}
\usepackage{epsf}
\title{ {\bf The effects of non-universal extra dimensions on the radiative
lepton flavor decays $\mu\rightarrow e\gamma$ and $\tau\rightarrow
\mu\gamma$ in the two Higgs doublet model.}}
\author{\vspace{1cm}\\
        {\bf E. O. Iltan}
        \thanks{E-mail address:
        eiltan@heraklit.physics.metu.edu.tr}
 \\
        Physics Department, Middle East Technical University \\
        Ankara, Turkey\\}

\date{}

\begin{document}
\setlength{\baselineskip}{24pt}
\maketitle
\setlength{\baselineskip}{7mm}
\begin{abstract}
We study the effect of non-universal extra dimensions on the
branching ratios of the lepton flavor violating  processes
$\mu\rightarrow e\gamma$ and $\tau\rightarrow \mu\gamma$ in the
general two Higgs doublet model. We observe that these effects are
small for a single extra dimension, however, in the case of two
extra dimensions there is a considerable enhancement in the
additional contributions.
\end{abstract}
\thispagestyle{empty}
\newpage
\setcounter{page}{1}
\section{Introduction}
Lepton flavor violating (LFV) interactions deserve to analyze
since, theoretically, they are sensitive the physics beyond the
standard model (SM) and they are rich in the sense that they exist
at loop level and make it possible to predict the free parameters
of the underlying theory.

The radiative LFV decays are among the well known candidates of
the LFV processes and there are various experimental and
theoretical works in the literature. The current limits for the
branching ratios ($BR$'s) of the $\mu\rightarrow e\gamma$ and
$\tau\rightarrow \mu\gamma$ decays are  $1.2\times 10^{-11}$
\cite{Brooks} and $1.1\times 10^{-6}$ \cite{Ahmed}, respectively.
The extensive theoretical work has been done on these decays in
the supersymmetric models \cite{Barbieri1}, in the framework of
the two Higgs doublet model (2HDM) \cite{Iltan1, Diaz} and also in
a model independent way \cite{Chang}.

In the general 2HDM, so called the model III, the possibility of
the flavor changing neutral currents (FCNC's) at tree level allows
the existence of the radiative LFV decays, theoretically. These
decays are induced by the internal neutral Higgs bosons $h^0$ and
$A^0$ and the Yukawa couplings carrying information about the
strength of the lepton-lepton-neutral Higgs interactions are free
parameters of the theory. These free parameters should be
restricted by using the experimental measurements.

This work is devoted to the LFV processes $\mu\rightarrow e\gamma$
and $\tau\rightarrow \mu\gamma$ in the framework of the model III,
with the inclusion of one (two) non-universal extra spatial
dimension. The extra dimension scenario is based on the string
theories as a possible solution to the hierarchy problem of the
standard model (SM). The effects of extra dimensions on various
phenomena have been studied in the literature
\cite{Arkani}-\cite{Lam}. For the extra dimension scenarios the
starting point is a fundamental theory, lying in higher dimensions
and the assumption that the four dimensional SM is its low energy
effective theory. The procedure to pass from higher dimensions to
the four dimension is the compactificitation of each extra
dimension to a circle $S^1$ with radius $R$, which is a typical
size of corresponding extra dimension. Finally, it leads to appear
new particles, namely Kaluza-Klein (KK) modes of the particles in
the theory. If all the fields feel the extra dimensions, so called
universal extra dimensions (UED), the extra dimensional momentum
is conserved, in other words, the KK number at each vertex is not
violated. In the case of a single UED, the compactification size
$R$ has been predicted as large as few hundereds of GeV
\cite{Arkani, Antoniadis1,Antoniadis3}, in the range $200-500\,
GeV$, using electroweak precision measurements \cite{Appelquist},
the \( B-\bar{B} \) -mixing \cite{Papavassiliou,Chakraverty} and
the flavor changing process $b \to s \gamma$ \cite{Agashe}. If the
extra dimensions are accessible to some fields but not all in the
theory, they are called non-universal extra dimensions. In this
case, the KK number at each vertex is not conserved and tree level
interaction of KK modes with the ordinary particles can exist. In
\cite{Dienes}, it was observed that, a very light left handed
neutrino can be obtained in the non-universal extra dimensions
where only the right handed neutrino feels the extra dimension.
The study in \cite{Agulia} is due to the effect of brane kinetic
terms for bulk scalars, fermions and gauge fields in higher
dimensional theories. There are also some phenomenological studies
on the non-universal extra dimensions in the literature
\cite{iltanEDM}.

In the present work, we study the branching ratios (BR's) of the
LFV processes $\mu\rightarrow e\gamma$ and $\tau\rightarrow
\mu\gamma$ in the model III and we assume that the extra
dimensions are felt by the new Higgs doublet and the gauge sector
of the theory. The compactification of the single (double) extra
dimension to the circle (torus) $S^1$ ($S^1\times S^1$) brings new
particles, KK modes of neutral Higgs bosons $h^0$ and $A^0$, and
the additional vertices, inducing KK mode of neutral Higgs
boson-lepton-lepton interaction, appear. In this case, the KK
number in the vertices is not conserved, in contrast to the
universal extra dimension (UED) case, where all fields experience
the extra dimensions. The non-zero KK modes of neutral Higgs
fields $H$ have masses $\sqrt{m_{H}^2+m_n^2}$
($\sqrt{m_{H}^2+m_n^2+m_r^2}$) with $m_n=n/R$ ($m_n=n/R$ ,
$m_r=r/R$). Here $m_n$ ($m_n^2+m_r^2$) is the mass (mass$^2$)of
$n$'th ($n,r$'th) level KK particle for a single (double)
non-universal extra dimension. Since the leptons have not KK modes
and the additional effects due to the extra dimensions are the
functions of the ratio $m^2_{lepton}/m^2_{KK\, neutral\, Higgs}$,
there is a suppression in the numerical values of the expressions
compared with the ones existing in the case of UED. However, we
observe that, the BR's of the processes under consideration
enhances considerably large for two non-universal extra
dimensions, due to the crowd of neutral Higgs KK modes.

The paper is organized as follows: In Section 2, we present the
BR's of LFV interactions $\mu\rightarrow e\gamma$ and
$\tau\rightarrow \mu\gamma$ in the model III version of the 2HDM
with the inclusion of non-universal extra dimensions. Section 3 is
devoted to discussion and our conclusions.
\section{The effects of non-universal extra dimensions on the radiative
lepton flavor decays $\mu\rightarrow e\gamma$ and $\tau\rightarrow
\mu\gamma$ in the two Higgs doublet model.}
The radiative LFV decays exist at least in the one loop level and
the theoretical values of their BR's enhance with the extension of
the Higgs sector in the models beyond the SM. The model III
version of the 2HDM accepts the flavor changing neutral currents
(FCNC) at tree level and it is the candidate model to bring
enhancement to the BR's of the radiative LFV decays. With the
inclusion of possible non-universal extra dimensions which are
felt by gauge bosons and the new Higgs particles, there appears an
additional contribution to the BR of the decays under
consideration. The part of Lagrangian which carries the
interaction responsible for the radiative LFV proceses in 5 (6)
dimension is
\begin{eqnarray}
{\cal{L}}_{Y}=
\xi^{E}_{5 (6)\, ij} \bar{l}_{i} (\phi_{2}|_{y (z)=0}) E_{j} +
h.c. \,\,\, . \label{lagrangian}
\end{eqnarray}
Here
$\xi^{E}_{5 (6)\, ij}$, with lepton family indices $i,j$, are $5
(6)$-dimensional dimensionful Yukawa couplings which can be
rescaled to the ones in 4-dimension as
$\xi^{E}_{5 (6)\, ij}=\sqrt{2 \pi R}\,(2 \pi R)\, \xi^{E}_{ij}$,
$\phi_{i}$ for $i=1,2$, are the two scalar doublets, $l_{i}$ and
$E_{j}$ are lepton doublets and singlets, respectively. The scalar
and lepton doublets are the functions of $x^\mu$ and $y$ ($y$,
$z$), where $y$ ($y$, $z$) is the coordinate represents the $5
(6)$'th dimension. With the choice of the doublets $\phi_{1}$ and
$\phi_{2}$
\begin{eqnarray}
\phi_{1}=\frac{1}{\sqrt{2}}\left[\left(\begin{array}{c c}
0\\v+H^{0}\end{array}\right)\; + \left(\begin{array}{c c}
\sqrt{2} \chi^{+}\\ i \chi^{0}\end{array}\right) \right]\, ;
\phi_{2}=\frac{1}{\sqrt{2}}\left(\begin{array}{c c}
\sqrt{2} H^{+}\\ H_1+i H_2 \end{array}\right) \,\, ,
\label{choice}
\end{eqnarray}
and the vacuum expectation values
\begin{eqnarray}
<\phi_{1}>=\frac{1}{\sqrt{2}}\left(\begin{array}{c c}
0\\v\end{array}\right) \,  \, ;
<\phi_{2}>=0 \,\, .
\label{choice2}
\end{eqnarray}
it is possible to switch off the mixing between neutral scalar
Higgs bosons and to separate the particle spectrum so that the SM
particles are collected in the first doublet and the new particles
in the second one. \footnote{Here $H^1$ ($H^2$) is the well known
mass eigenstate $h^0$ ($A^0$).} Notice that the Yukawa matrices
$\xi^{E}_{5 (6)\, ij}$ are responsible for producing the FCNC at
tree level and their entries are complex in general. Notice that,
in the following, we replace $\xi^{E}$ with $\xi^{E}_{N}$ where
"N" denotes the word "neutral".

In the case of the non-universal two extra dimensions where only
the new Higgs field $\phi_{2}$ is accessible to extra dimension in
the Higgs sector, the compactification on a torus $S^1\times S^1$,
causes to appear KK modes $\phi_{2}^{(n,r)}$ of $\phi_{2}$ in two
spatial extra dimension as
\begin{eqnarray}
\phi_{2}(x,y,z ) & = & \frac{1}{(2 \pi R)^{d/2}} \left\{
\phi_{2}^{(0,0)}(x) + 2^{d/2} \sum_{n,r}^{\infty}
\phi_{2}^{(n,r)}(x) \cos(ny/R+rz/R)\right\} \,
\label{SecHiggsField}
\end{eqnarray}
where $d=2$, the indices  $n$ and $r$ are positive integers
including zero but both are not zero at the same time. Here
$\phi_{2}^{(0,0)}(x)$ the 4-dimensional Higgs doublet which
includes the charged Higgs boson $H^+$, the neutral CP even (odd)
$h^0$ ($A^0$) Higgs bosons. The KK mode of the charged Higgs boson
has the mass $\sqrt{m_{H^\pm}^2+m_n^2+m_r^2}$. Similarly, the
neutral CP even (odd) Higgs $h^0$ ($A^0$) has the mass
$\sqrt{m_{h^0}^2+m_n^2+m_r^2}$ ($\sqrt{m_{A^0}^2+m_n^2+m_r^2}$),
where $m_n=n/R$ and $m_r=r/R$. Furthermore, we assume that the
compactification radius $R$ is the same for both dimensions. The
expansion for a single extra dimension can be obtained by setting
$d=1$, taking $z=0$, and dropping the summation over $r$. Notice
that the gauge fields feel also the extra dimensions which results
in KK modes after the compactification. However, they are
irrelevant in our calculation since they do not play any role in
the processes under consideration.

\begin{figure}[htb]
\vskip 2.0truein \centering \epsfxsize=6.8in
\leavevmode\epsffile{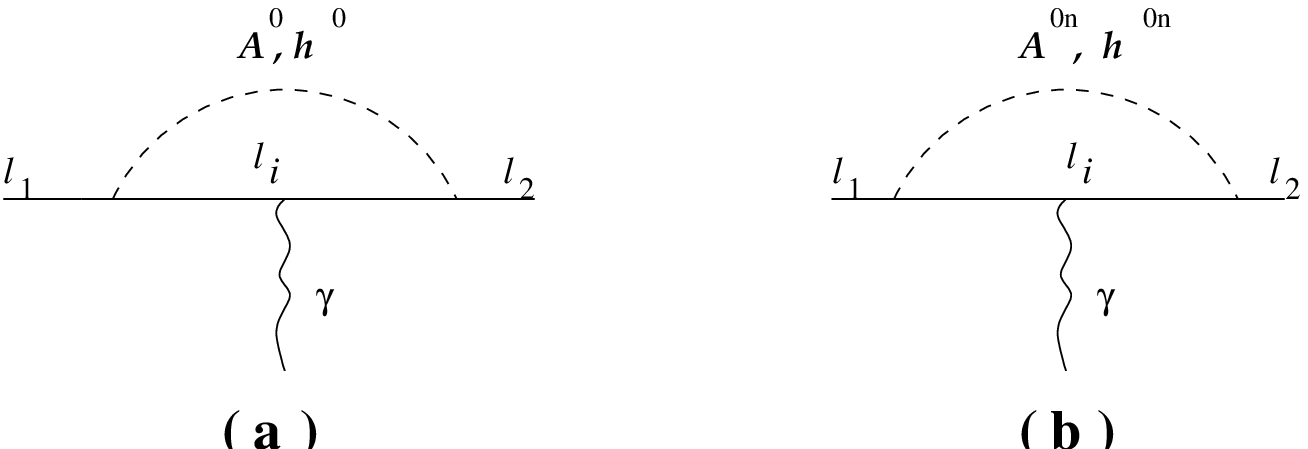} \vskip 1.0truein \caption[]{One loop
diagrams contribute to $l_1\rightarrow l_2 \gamma$ decay  due to
the zero mode (KK mode) neutral Higgs bosons $h^0$ and $A^0$
($h^{0 n}$ and $A^{0 n}$) in the 2HDM.} \label{fig1}
\end{figure}

At this stage, we consider the flavor violating radiative decays
of leptons,  $\mu\rightarrow e\gamma$ and $\tau\rightarrow
\mu\gamma$, by introducing a single non-universal extra spatial
dimension. These decays exist at least at the one loop level with
the help of the intermediate neutral Higgs bosons $h^0$ and $A^0$.
With the inclusion of a single non-universal extra dimension, the
KK modes of neutral Higgs bosons, namely $h^{0 n}$ and $A^{0 n}$,
contribute after the compactification (Fig. \ref{fig1}). Here the
vertices involve two zero modes and one KK mode, since the KK mode
conservation does not exist in the case of non-universal extra
dimension, in contrary to the universal one.

In the loop calculations, the logarithmic divergences are
eliminated by using the on-shell renormalization scheme. In this
scheme, the self energy diagrams for on-shell leptons vanish since
they can be written as
\begin{eqnarray}
\sum(p)=(\hat{p}-m_{l_1})\, \bar{\sum}(p)\, (\hat{p}-m_{l_2})\,\,
, \label{self}
\end{eqnarray}
The divergences coming from the vertex diagrams can be eliminated
by introducing a counter term $V^{C}_{\mu}$ so that we have the
relation
\begin{eqnarray}
V^{Ren}_{\mu}=V^{0}_{\mu}+V^{C}_{\mu} \,\, , \label{Vert}
\end{eqnarray}
with the renormalized (bare) vertex $V^{Ren}_{\mu}$
($V^{0}_{\mu}$) and using the gauge invariance:
\begin{eqnarray}
k^{\mu} V^{Ren}_{\mu}=0 \,\, , \label{Vert2}
\end{eqnarray}
where $k^{\mu}$ is the photon four momentum vector.

Taking the vertex diagrams $a$ and $b$ in Fig. \ref{fig1} and
introducing only $\tau$ lepton for the internal line, the decay
width $\Gamma$ reads as
\begin{eqnarray}
\Gamma (\mu\rightarrow e\gamma)=c_1(|A_1|^2+|A_2|^2)\,\, ,
\label{DWmuegam}
\end{eqnarray}
where
\begin{eqnarray}
A_1&=&Q_{\tau} \frac{1}{48\,m_{\tau}^2} \Bigg (6\,m_\tau\,
\bar{\xi}^{E *}_{N,\tau e}\, \bar{\xi}^{E *}_{N,\tau \mu}\, \Big(
(F (z_{h^0})-F (z_{A^0}))+2\, \sum_{n=1}^{\infty} (F (z_{n,
h^0})-F (z_{n, A^0}))\Big ) \nonumber \\ &+& m_{\mu}\,\bar{\xi}^{E
*}_{N,\tau e}\, \bar{\xi}^{E}_{N,\tau \mu}\, \Big(G (z_{h^0})+G
(z_{A^0})+2\,\sum_{n=1}^{\infty} (G (z_{n, h^0})+G (z_{n,
A^0}))\Big) \Bigg)
\nonumber \,\, , \\
A_2&=&Q_{\tau} \frac{1}{48\,m_{\tau}^2} \Bigg (6\,m_\tau\,
\bar{\xi}^{D}_{N,\tau e}\, \bar{\xi}^{D}_{N,\tau \mu}\, \Big( (F
(z_{h^0})-F (z_{A^0}))+2\,\sum_{n=1}^{\infty} (F (z_{n, h^0})-F
(z_{n, A^0}))\Big )\nonumber \\ &+& m_{\mu}\,\bar{\xi}^{D}_{N,\tau
e}\, \bar{\xi}^{D *}_{N,\tau \mu}\, \Big( G (z_{h^0})+G (z_{A^0})+
2\,\sum_{n=1}^{\infty} (G (z_{n, h^0})+ G (z_{n, A^0})) \Big)
\Bigg)
 \,\, , \label{A1A2}
\end{eqnarray}
with $c_1=\frac{G_F^2 \alpha_{em} m^3_{\mu}}{32 \pi^4}$, left
(right) chiral amplitudes $A_1$ ($A_2$). Similarly, the decay
width of the LFV process $\tau\rightarrow \mu\gamma$ can be
obtained as
\begin{eqnarray}
\Gamma (\tau\rightarrow \mu\gamma)=c_2(|B_1|^2+|B_2|^2)\,\, ,
\end{eqnarray}
\label{DWtaumugam}
where
\begin{eqnarray}
B_1&=&Q_{\tau} \frac{1}{48\,m_\tau^2} \Bigg (6\,m_{\tau}\,
\bar{\xi}^{E *}_{N,\tau \mu}\, \bar{\xi}^{E *}_{N,\tau \tau}\,
\Big( (F (z_{h^0})-F (z_{A^0}))+2\,\sum_{n=1}^{\infty} (F (z_{n,
h^0})-F (z_{n, A^0}))\Big) \nonumber \\ &+& m_{\tau}\,\bar{\xi}^{E
*}_{N,\tau \mu}\, \bar{\xi}^{E}_{N,\tau \tau}\, \Big( G
(z_{h^0})+G (z_{A^0})+2\,\sum_{n=1}^{\infty} (G (z_{n, h^0})+ G
(z_{n, A^0}))\Big)  \Bigg)
\nonumber \,\, , \\
B_2&=&Q_{\tau} \frac{1}{48\,m_\tau^2} \Bigg (6\,m_{\tau}\,
\bar{\xi}^{D}_{N,\tau \mu}\, \bar{\xi}^{D}_{N,\tau \tau}\, \Big(
(F(z_{h^0})-F (z_{A^0}))+2\,\sum_{n=1}^{\infty} (F (z_{n, h^0})-F
(z_{n, A^0})) \Big)\nonumber \\ &+&
m_{\tau}\,\bar{\xi}^{D}_{N,\tau \mu}\, \bar{\xi}^{D *}_{N,\tau
\tau}\, \Big( G (z_{h^0})+G (z_{A^0})+2\,\sum_{n=1}^{\infty} (G
(z_{n, h^0})+G (z_{n, A^0}))\Big)  \Bigg)
 \,\, , \label{B1B2}
\end{eqnarray}
with $c_1=\frac{G_F^2 \alpha_{em} m^3_{\mu}}{32 \pi^4}$, left
(right) chiral amplitudes $B_1$ ($B_2$).  The functions $F (w)$
and $G (w)$ in eqs. (\ref{A1A2}) and (\ref{B1B2}) read
\begin{eqnarray}
F (w)&=&\frac{w\,(3-4\,w+w^2+2\,ln\,w)}{(-1+w)^3} \, , \nonumber \\
G (w)&=&\frac{w\,(2+3\,w-6\,w^2+w^3+ 6\,w\,ln\,w)}{(-1+w)^4} \,\,
, \label{functions2}
\end{eqnarray}
and $z_{H}=\frac{m^2_{\tau}}{m^2_{H}}$, $z_{n,
H}=\frac{m^2_{\tau}}{m^2_{H}+(n/R)^2}$, $Q_{\tau}$ is the charge
of $\tau$ lepton, the dimensionfull Yukawa couplings
$\bar{\xi}^{E}_{N,ij}$ appearing in the expressions are defined as
$\xi^{E}_{N,ij}=\sqrt{\frac{4\,G_F}{\sqrt{2}}}\,
\bar{\xi}^{E}_{N,ij}$. Furthermore, the theoretical results is due
to the internal $\tau$-lepton contribution since, we respect the
Sher  scenerio \cite{Sher}, results in the couplings
$\bar{\xi}^{E}_{N, ij}$ ($i,j=e,\mu$), are small compared to
$\bar{\xi}^{E}_{N,\tau\, i}$ $(i=e,\mu,\tau)$, due to the possible
proportionality of them to the masses of leptons under
consideration in the vertices. Furthermore, we take the Yukawa
couplings for the interactions lepton-lepton-KK mode of Higgs
bosons ($h^0$ and $A^0$) as the same as the ones for the
interactions of zero mode fields.

In the case two extra spatial dimensions which are felt by the
second Higgs doublet in the Higgs sector, $\phi_{2}$ can be
expanded into its KK modes as in eq. (\ref{SecHiggsField}) after
the compactification of the extra dimensions  on a torus
$S^1\times S^1$. In the two non-universal spatial extra dimensions
the forms  of the expressions (\ref{A1A2}) and (\ref{B1B2}) are
the same, except that, the parameters $z_{n, H}$ and
$\sqrt{m_{H}^2+m_n^2}$ are replaced by , $z_{n,r,
H}=\frac{m^2_{\tau}}{m^2_{H}+(n/R)^2+(r/R)^2}$ and
$\sqrt{m_{H}^2+m_n^2+m_r^2}$. Furthermore, the number $2$ in front
of the summation should be replaced by $4$ and the summation
should be done for  $n,r=0,1,2 ...$ except $n=r=0$.
\section{Discussion}
The radiative LFV decays $\mu\rightarrow e\gamma$ and
$\tau\rightarrow \mu\gamma$ are controlled by the Yukawa couplings
$\bar{\xi}^D_{N,ij}, \, i,j=e, \mu, \tau$  in the model III
version of the 2HDM. These couplings should be restricted by using
the experimental results, since they are free parameters of the
theory. The dominant couplings are $\bar{\xi}^{E}_{N,\tau i}$,
($i=e, \mu$) and $\bar{\xi}^{E}_{N,\tau i}$ ($i=\mu, \tau$) for
the $\mu\rightarrow e\gamma$ and $\tau\rightarrow \mu\gamma$
decays, respectively. Here, we assume that, the strength of these
couplings are related with the masses of leptons denoted by the
indices of them, similar to the Cheng-Sher scenerio \cite{Sher}.
This assumption forces us to consider only the $\tau$ lepton at
the internal line for both processes. We also assume that the
Yukawa couplings $\bar{\xi}^{E}_{N,ij}$ is symmetric with respect
to the indices $i$ and $j$. For the numerical values of these
couplings we use the following restrictions:
\begin{itemize}
\item The upper limit of $\bar{\xi}^{E}_{N,\tau \mu}$ is predicted
as $30\, GeV$ (see \cite{Iltananomuon} and references therein) by
using the experimental uncertainty, $10^{-9}$, in the measurement
of the muon anomalous magnetic moment and assuming that the new
physics effects can not exceed this uncertainty.
\item For the Yukawa coupling $\bar{\xi}^{E}_{N,\tau e}$, we use
broad range of the prediction, $10^{-3}-10^{-2}\, GeV$, which is
obtained by using the experimental upper limit of BR of
$\mu\rightarrow e \gamma$ decay, $BR\leq 1.2\times 10^{-11}$ and
predicted upper limit of $\bar{\xi}^{E}_{N,\tau \mu}\leq 30\, GeV$
(see \cite{Iltan1}).

\item For the Yukawa couplings $\bar{\xi}^{E}_{N,\tau \tau}$, we
use the numerical values which is greater than the upper limit of
$\bar{\xi}^{E}_{N,\tau \mu}$.
\end{itemize}

With the addition of the extra dimensions, the new contributions
to the physical parameters arise for these decays. In the case of
non-universal extra dimensions, the KK number at each vertex is
not conserved in contrast to the universal one case and
lepton-lepton-KK neutral Higgs  vertices appear. Due to the
compactification of the a single (double) extra dimension on a
circle (torus) $S^1$ ($S^1\times S^1$), the KK modes of internal
neutral Higgs bosons $h^0$ and $A^0$ appear and they bring
additional contributions to the physical quantities of the decays
under consideration.

Our work is devoted to the prediction of the non-universal extra
dimensions on the BR of the LFV processes $\mu\rightarrow e
\gamma$ and $\tau \rightarrow \mu\gamma$, in the framework of the
type III 2HDM. We make this anaysis in one and two extra
dimensions. In the case of two extra dimensions the crowd of  KK
modes cause to be more enhancement in the $BR$ of the LFV decays
compared to the case of a single extra dimension. On the other
hand, the abundance of KK modes in the summation can lead to the
divergence in the calculation of the $BR$. However the the ratio
$\frac{m_{\tau}^2}{m_{H}^2+m_n^2+m_r^2}$ appearing in the
expressions converges to zero sharply with the increasing values
of the integers $n$ and $r$ and it leads to the convergence of the
KK sum for the compactification scale we study, $1/R > 200\, GeV$.
For the UED case the above ratio has the form
$\frac{m_{\tau}^2+m_n^2+m_r^2}{m_{H}^2+m_n^2+m_r^2}$ since KK
number at each vertex is conserved and KK lepton-lepton-KK neutral
Higgs interactions are switched on. Increasing values of $n$ and
$r$ forces the ratio to reach one and the convergence problem of
the KK sum should be studied carefully.

\begin{figure}[htb]
\vskip -3.0truein \centering \epsfxsize=6.8in
\leavevmode\epsffile{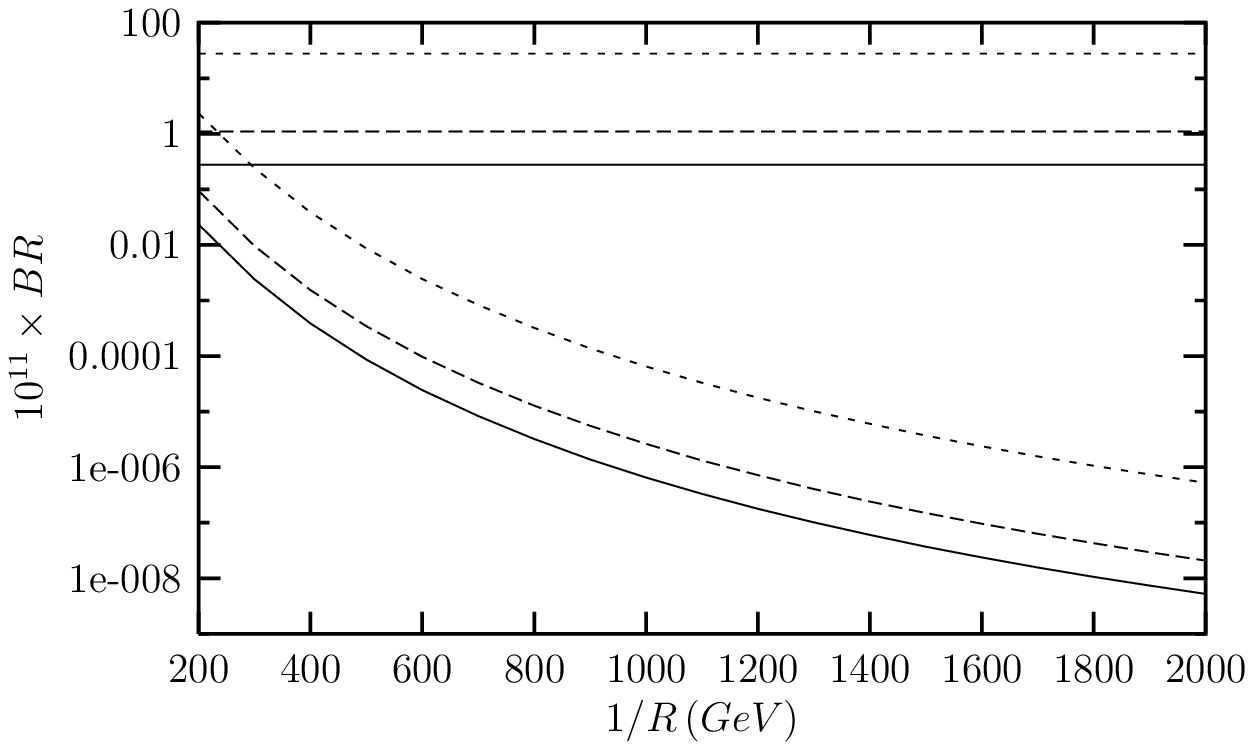} \vskip -3.0truein \caption[]{
The compactification scale $1/R$ dependence of the BR of the  LFV
decay $\mu\rightarrow e \gamma$ for $m_{h^0}=100\, GeV$,
$m_{A^0}=200\, GeV$, $\bar{\xi}^{D}_{N,\tau\mu}=30\, GeV$, and
four different values of the coupling $\bar{\xi}^{D}_{N,\tau e}$,
in the case of one non-universal extra dimension. The
solid-dashed-small dashed straight lines (curves) represent the
2HDM (the extra dimension) contribution to the BR for
$\bar{\xi}^{D}_{N,\tau e}=0.5\times 10^{-3}-1.0\times
10^{-3}-0.5\times 10^{-2}\,GeV$. } \label{BRmuegamR1}
\end{figure}

\begin{figure}[htb]
\vskip -3.0truein \centering \epsfxsize=6.8in
\leavevmode\epsffile{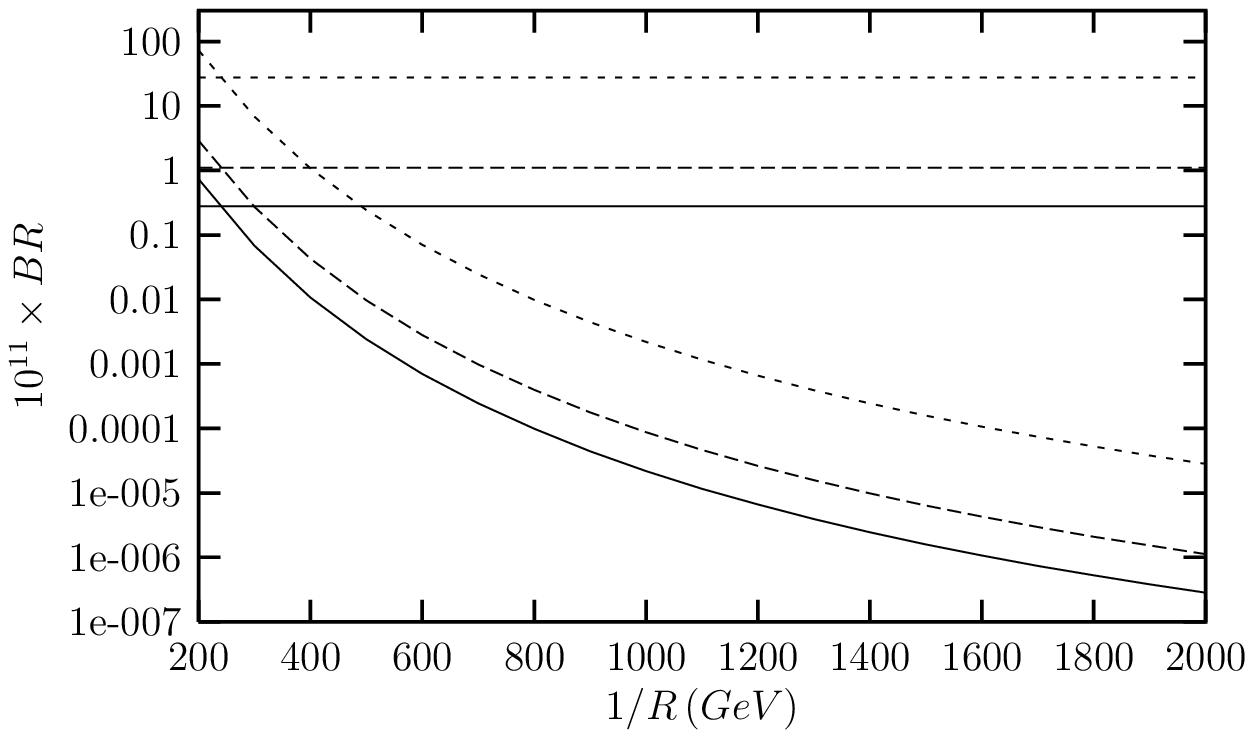} \vskip -3.0truein \caption[]{
The same as Fig. \ref{BRmuegamR1} but for two non-universal extra
dimensions.} \label{BRmuegamR2}
\end{figure}

Fig. \ref{BRmuegamR1} (\ref{BRmuegamR2}) is devoted to the
compactification scale $1/R$ dependence of the BR of the  LFV
decay $\mu\rightarrow e \gamma$, for $m_{h^0}=100\, GeV$,
$m_{A^0}=200\, GeV$, $\bar{\xi}^{D}_{N,\tau\mu}=30\, GeV$, and
four different values of the coupling $\bar{\xi}^{D}_{N,\tau e}$,
in the case of one (two) non-universal extra dimension(s). In Fig.
\ref{BRmuegamR1}, the solid-dashed-small dashed straight lines
(curves) represent the 2HDM (the extra dimension) contribution to
the BR for $\bar{\xi}^{D}_{N,\tau e}=0.5\times 10^{-3}-1.0\times
10^{-3}-0.5\times 10^{-2}\,GeV$. It is shown that the BR is not
sensitive to the extra dimension effects for the single extra
dimension. In the case of two non-universal extra dimensions (see
Fig. \ref{BRmuegamR2}), the sensitivity increases and the effects
of extra dimensions can reach even to the main contribution for
the small values of compactification scale. This enhancement in
the case of two extra dimensions is due to the abundance of KK
modes of neutral Higgs bosons.

\begin{figure}[htb]
\vskip -3.0truein \centering \epsfxsize=6.8in
\leavevmode\epsffile{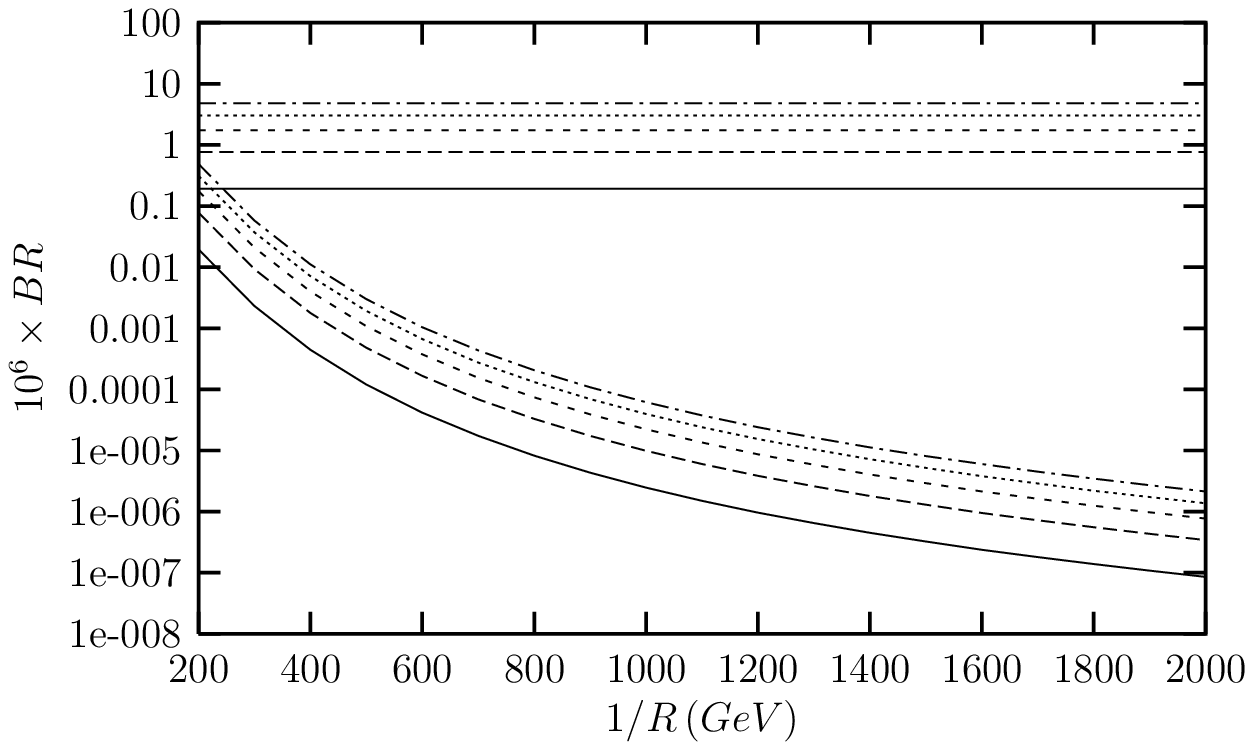} \vskip -3.0truein
\caption[]{ The compactification scale $1/R$ dependence of the BR
of the LFV decay $\tau\rightarrow \mu \gamma$ for $m_{h^0}=100\,
GeV$, $m_{A^0}=200\, GeV$, $\bar{\xi}^{D}_{N,\tau\mu}=30\, GeV$,
and five different values of the coupling $\bar{\xi}^{D}_{N,\tau
\tau}$, in the case of a single non-universal extra dimension. The
solid-dashed-small dashed-dotted-dot dashed straight lines
(curves) represent the 2HDM (the extra dimension) contribution to
the BR for $\bar{\xi}^{D}_{N,\tau e}=50-100-150-200-250\,GeV$.}
\label{BRtaumugamR1}
\end{figure}

\begin{figure}[htb]
\vskip -3.0truein \centering \epsfxsize=6.8in
\leavevmode\epsffile{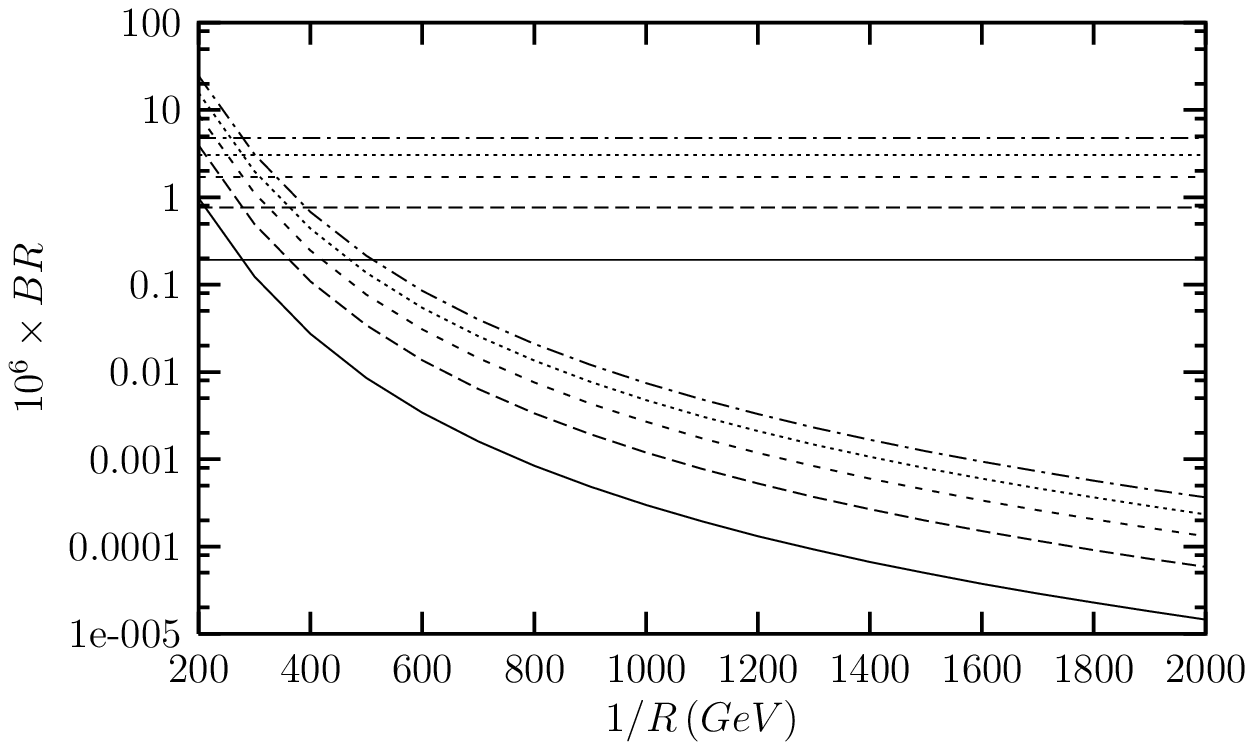} \vskip -3.0truein
\caption[]{The same as Fig. \ref{BRtaumugamR1} but for two
non-universal extra dimensions.} \label{BRtaumugamR2}
\end{figure}

In Fig. \ref{BRtaumugamR1} (\ref{BRtaumugamR2}) we present the
compactification scale $1/R$ dependence of the BR of the LFV decay
$\tau\rightarrow \mu \gamma$ for $m_{h^0}=100\, GeV$,
$m_{A^0}=200\, GeV$, $\bar{\xi}^{D}_{N,\tau\mu}=30\, GeV$, and
five different values of the coupling $\bar{\xi}^{D}_{N,\tau
\tau}$, in the case of one (two) non-universal extra dimension(s).
In Fig. \ref{BRtaumugamR1} the solid-dashed-small
dashed-dotted-dot dashed straight lines (curves) represent the
2HDM (the extra dimension) contribution to the BR for
$\bar{\xi}^{D}_{N,\tau e}=50-100-150-200-250\,GeV$. Similar to the
$\mu\rightarrow e \gamma$ decay the BR is not sensitive to the
extra dimension effects for the single extra dimension. However
Fig. \ref{BRtaumugamR2} shows that the BR is much more sensitive
to the contribution coming from the KK modes in the case of two
exta dimensions, in comparison with the single extra one. This
sensitivity becomes weak for $1/R \geq 600 \, GeV$.

Now we would like to analyze the ratio
$Ratio=\frac{BR^{Ext}}{BR^{Exp}-BR^{2HDM}}$ where $BR^{Exp}$
($BR^{2HDM}$) is the experimental upper limit (theoretical value)
of the $BR(\tau\rightarrow \mu\gamma)$ and $BR^{Ext}$ is the new
contributions coming from the extra dimensions.

\begin{figure}[htb]
\vskip -3.0truein \centering \epsfxsize=6.8in
\leavevmode\epsffile{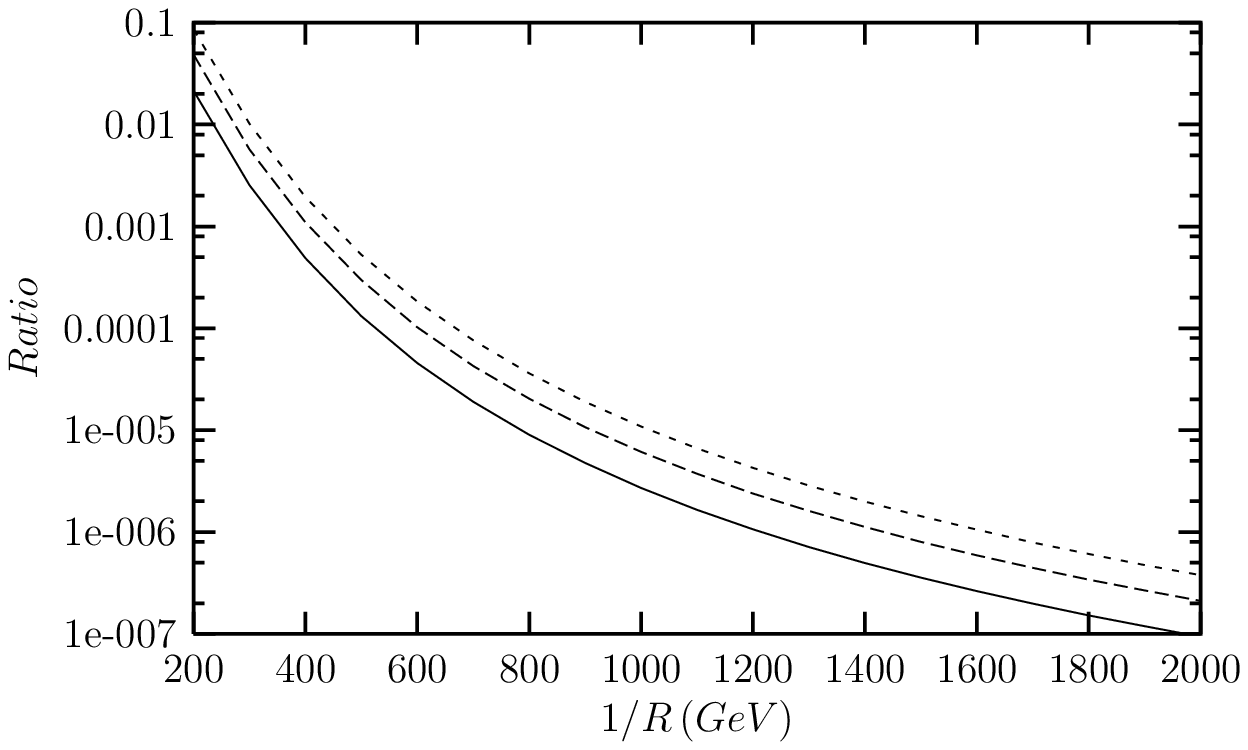} \vskip -3.0truein
\caption[]{The compactification scale $1/R$ dependence of the
$Ratio$ for the LFV decay $\tau\rightarrow \mu \gamma$ for
$m_{h^0}=100\, GeV$, $m_{A^0}=200\, GeV$,
$\bar{\xi}^{D}_{N,\tau\mu}=30\, GeV$, and three different values
of the coupling $\bar{\xi}^{D}_{N,\tau \tau}$, in the case of one
non-universal extra dimension. The solid-dashed-small dashed lines
represent the $Ratio$ for $\bar{\xi}^{D}_{N,\tau
e}=50-75-100\,GeV$} \label{Ratiotaumugam1}
\end{figure}

\begin{figure}[htb]
\vskip -3.0truein \centering \epsfxsize=6.8in
\leavevmode\epsffile{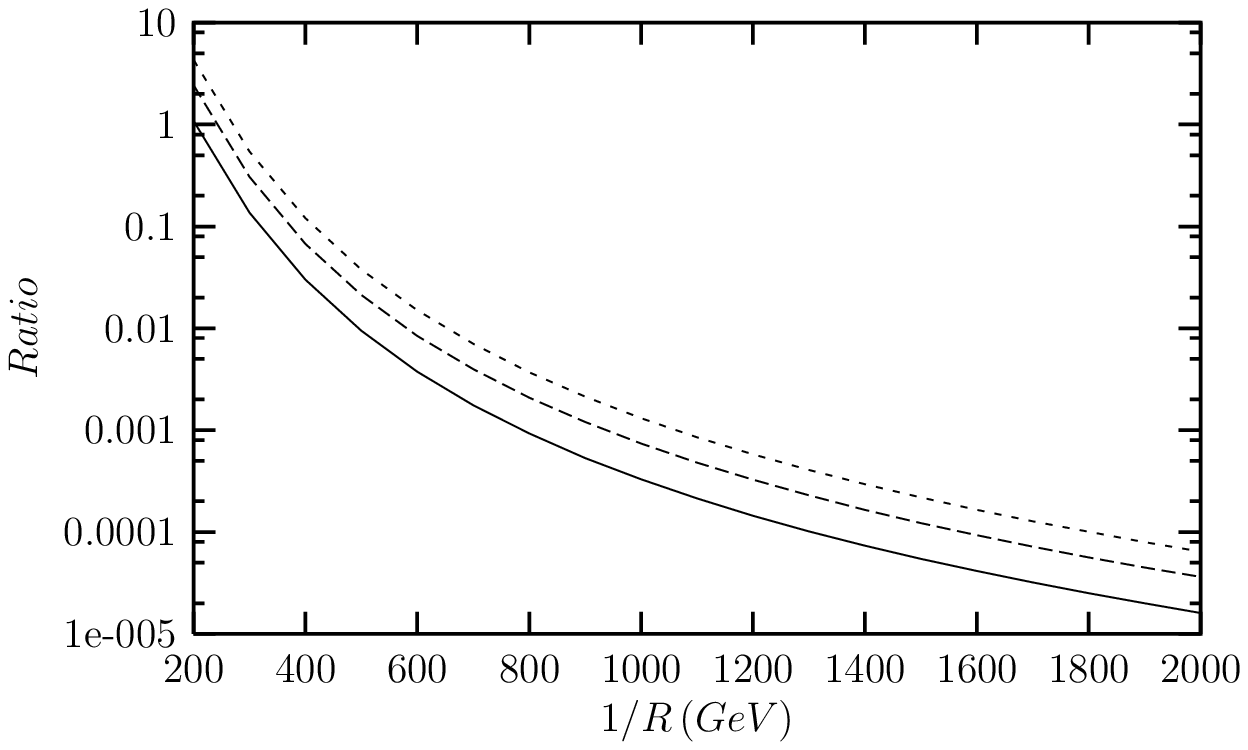} \vskip -3.0truein
\caption[]{The same as Fig. \ref{Ratiotaumugam1} but for two
non-universal extra dimensions.} \label{Ratiotaumugam2}
\end{figure}

Fig. \ref{Ratiotaumugam1} (\ref{Ratiotaumugam2}) represents  the
compactification scale $1/R$ dependence of the $Ratio$ for the LFV
decay $\tau\rightarrow \mu \gamma$ for $m_{h^0}=100\, GeV$,
$m_{A^0}=200\, GeV$, $\bar{\xi}^{D}_{N,\tau\mu}=30\, GeV$, and
three different values of the coupling $\bar{\xi}^{D}_{N,\tau
\tau}$ in the case of one (two) non-universal extra dimension(s).
In Fig. \ref{Ratiotaumugam1} (\ref{Ratiotaumugam2}) the
solid-dashed-small dashed lines represent the $Ratio$ for
$\bar{\xi}^{D}_{N,\tau e}=50-75-100\,GeV$, for a single (double)
non-universal extra dimension. Fig. \ref{Ratiotaumugam1} shows
that the discrepancy between the experimental value and the 2HDM
prediction cannot be compensated with the addition of extra
dimensions. On the other hand this ratio increases with the
increasing values of the coupling constant
$\bar{\xi}^{D}_{N,\tau\tau}$. For two non-universal extra
dimensions it is possible to cover the discrepancy between the
experimental and the theoretical values for the small values of
the compactification scale $1/R$. For $600\leq 1/R \leq 800/, GeV$
the magnitude of extra dimension effect becomes almost $1\%$ of
the difference of the experimental and the theoretical values of
the BR.

\begin{figure}[htb]
\vskip -3.0truein \centering \epsfxsize=6.8in
\leavevmode\epsffile{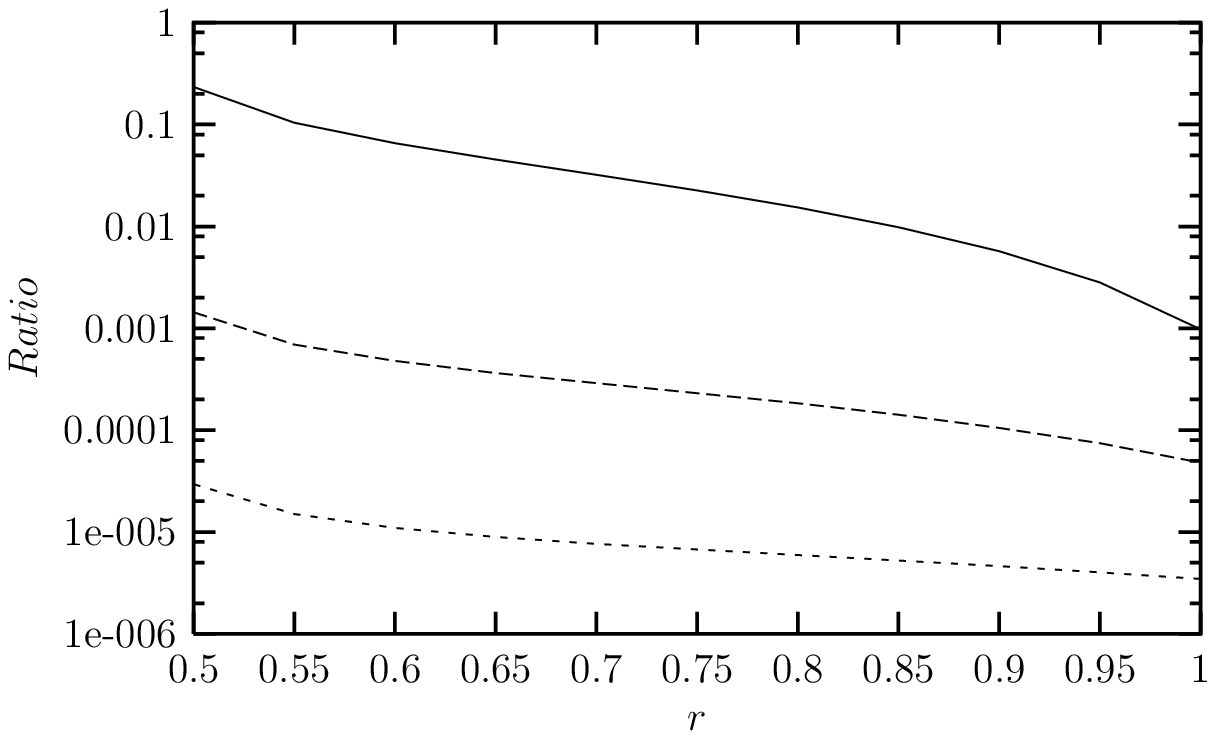} \vskip -3.0truein
\caption[]{The parameter $r=\frac{m_h^0}{m_A^0}$ dependence  of
the $Ratio$ for the $\tau\rightarrow \mu \gamma$ for
$m_{A^0}=200\, GeV$, $\bar{\xi}^{D}_{N,\tau\mu}=30\, GeV$,
$\bar{\xi}^{D}_{N,\tau \tau}=100\, GeV$, for three different
values of the compactification scale $1/R$, $1/R=200, 500, 1000 \,
GeV$ in the case of one non-universal extra dimension.}
\label{Ratiotaumugamrr1}
\end{figure}

\begin{figure}[htb]
\vskip -3.0truein \centering \epsfxsize=6.8in
\leavevmode\epsffile{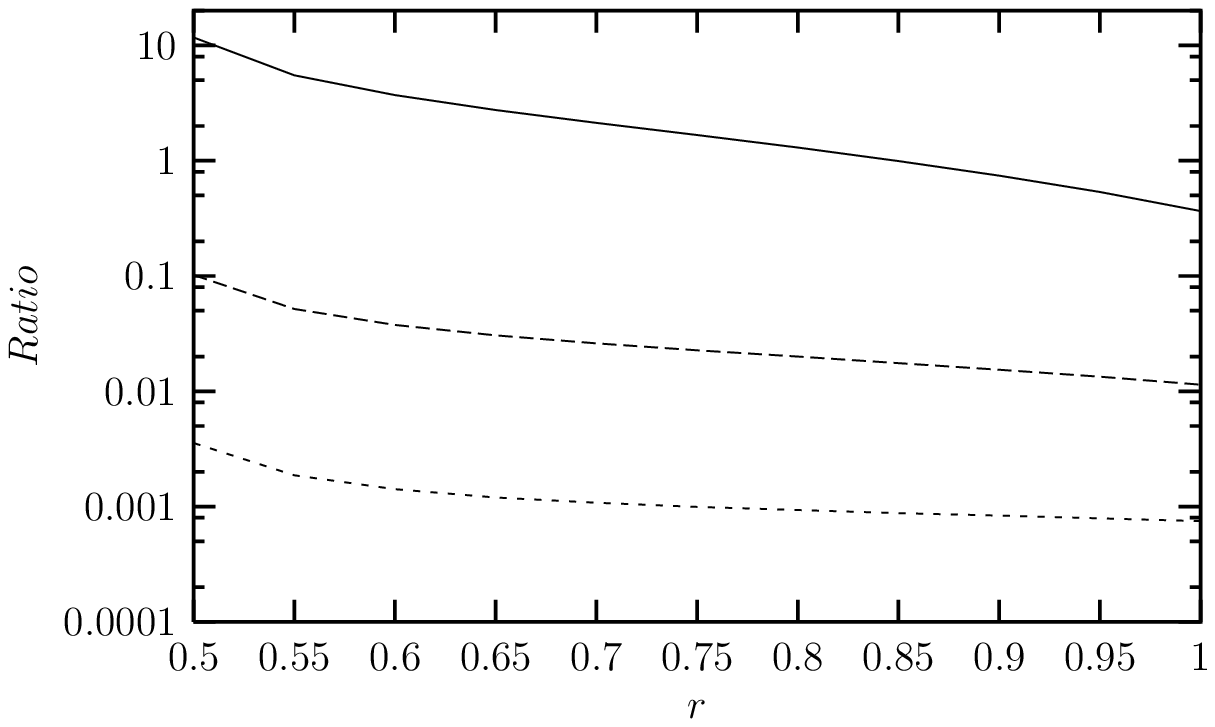} \vskip -3.0truein
\caption[]{The same as Fig. \ref{Ratiotaumugamrr1} but for two
non-universal extra dimensions.} \label{Ratiotaumugamrr2}
\end{figure}

In Fig. \ref{Ratiotaumugamrr1} (\ref{Ratiotaumugamrr2}) we present
the parameter $r=\frac{m_{h^0}}{m_{A^0}}$ dependence  of the
$Ratio$ for the $\tau\rightarrow \mu \gamma$ decay for
$m_{A^0}=200\, GeV$, $\bar{\xi}^{D}_{N,\tau\mu}=30\, GeV$,
$\bar{\xi}^{D}_{N,\tau \tau}=100\, GeV$, and three different
values of the compactification scale $1/R$, $1/R=200, 500, 1000 \,
GeV$ in the case of one (two) non-universal extra dimension(s).
These figures show that the $Ratio$ is sensitive to the mass ratio
of neutral Higgs bosons $h^0$ and $A^0$ especially for small
$1/R$. The increasing values of $r$ causes to decrease the ratio
and this sensitivity becomes weak for the large values of the
scale $1/R$. In the case that the neutral Higgs masses are nearly
degenerate the $Ratio$ is suppressed almost one order compared to
the case that the parameter $r=0.5$ for the intermediate values of
$1/R$.

Now we would like to present the results briefly.
\begin{itemize}
\item The single non-universal extra dimension effects on the BR's
of $\mu\rightarrow e \gamma$ and $\tau\rightarrow \mu \gamma$
decays are considerably weak, however, the addition of one more
spatial dimension causes relatively strong sensitivity on the
BR's.
\item  The contribution coming from the single non-universal extra
dimension is not enough large to compensate the difference between
the experimental and the theoretical values of the BR. In the case
of two non-universal extra dimensions this discrepancy can be
covered by KK modes of neutral Higgs bosons for the small values
of the compactification scale $1/R$.
\item The extra dimension contributions are sensitive to the mass
ratio of neutral Higgs bosons $h^0$ and $A^0$ and it becomes
weaker with the increasing values of the compactification scale
$1/R$.
\end{itemize}
As a final comment, the effect of the single non-universal extra
dimension on the BR's of LFV decays $\mu\rightarrow e\gamma$ and
$\tau\rightarrow \mu\gamma$ is weak. However, their contributions
enhance in the case of two extra dimensions and the more accurate
future experimental results of these decays, hopefully, will be
helpful to analyze the possible signals coming from the extra
dimensions.
\section{Acknowledgement}
This work has been supported by the Turkish Academy of Sciences in
the framework of the Young Scientist Award Program.
(EOI-TUBA-GEBIP/2001-1-8)
\newpage
\newpage
\end{document}